\documentclass[11pt]{article}
\usepackage{amsmath}

\usepackage{EACL2023}

\usepackage{times}
\usepackage{latexsym}

\usepackage[T1]{fontenc}

\usepackage[utf8]{inputenc}

\usepackage{microtype}

\usepackage{inconsolata}
\usepackage{graphicx}
\usepackage{booktabs}
\graphicspath{ {images/} }

\title{Scalable Audio-LLM System for Speech MOS Evaluation: A Diagnostic Case Study and
Adaptation Roadmap}

\author{
  \parbox{0.9\textwidth}{
    \centering
    Dimitrios Kokkinis\textsuperscript{1}, 
    Panagiotis Kakoulidis\textsuperscript{1}, 
    Nikolaos Ellinas\textsuperscript{1}, 
    Chrysovalanti Lymperi\textsuperscript{1},
    Alexandra Vioni\textsuperscript{1}, 
    Georgios Vardaxoglou\textsuperscript{1}, 
    Iakovi Alexiou\textsuperscript{1}, 
    Myrsini Christidou\textsuperscript{1}, 
    Jeasung Lee\textsuperscript{2}, 
    Gunu Jho\textsuperscript{2}, 
    Inchul Hwang\textsuperscript{2}, 
    Pirros Tsiakoulis\textsuperscript{1}, 
    Aimilios Chalamandaris\textsuperscript{1} \\[1em]
    \textsuperscript{1}\normalfont{Innoetics, Samsung Electronics, Greece} \\ 
    \textsuperscript{2}Mobile eXperience Business, Samsung Electronics, Republic of Korea
  }
}  

\begin{document}
\maketitle
\begin{abstract}

The continuous deployment of enterprise voice systems creates a critical operational 
bottleneck: evaluating the acoustic quality of synthetic speech across high-volume production 
updates. While human Mean Opinion Score (MOS) remains the quality gold standard, it 
introduces severe operational costs and turn-around delays that make continuous integration infeasible. 
In this work, we present a system design case study for a scalable, low-cost and offline 
evaluation pipeline capable of scoring large batches of synthetic speech at marginal cost. 
First, we use Qwen2-Audio-7B-Instruct in a zero-shot setting for granular MOS estimation, 
noticing perceptual blind spots and inherent biases. Then, we analyze SOMOS dataset and 
construct a diagnostic benchmark that operationalizes two prosodic dimensions associated with 
human naturalness ratings. Finally, we adapt Qwen2-Audio-7B-Instruct through a 
parameter-efficient finetuning (LoRA) on the task of MOS evaluation, observing improvements 
over base model and widely used neural baselines like UTMOS and establishing a practical 
blueprint for enterprise-grade speech quality assurance.

\end{abstract}

\section{Introduction}

Modern enterprise voice pipelines such as Text-to-Speech (TTS) systems \citep{qwen3tts}
\cite{fishaudios2} undergo continuous model updates, fine-tuning, and prompt engineering in 
modern CI/CD setups. Maintaining consistent perceptual quality requires rigorous quality 
assurance prior to production deployment. Historically, human MOS 
has served as the gold standard for speech quality evaluation. However, relying on 
human-in-the-loop testing for frequent model updates and thousands of generated audio 
files, introduces severe operational bottlenecks: they are prohibitively expensive, suffer 
from multi-day turnaround latencies, and cannot scale to evaluate large volumes of offline 
audio files. On the other hand, traditional regression-based MOS predictors are not reliable 
enough to act as the final quality assurance. Consequently, there is an urgent industry need 
for a lightweight, automated quality evaluation system capable of predicting speech quality.

LLM-as-a-Judge has been widely adopted in the text-domain, serving as a high-quality
alternative to human perception, where the model can confidently provide a thorough
quality evaluation to LLM-generated content. The question remains, whether Audio-LLMs can
be used in a speech evaluation system effectively. To examine this, we first conduct a baseline feasibility 
assessment using Qwen2-Audio-7B-Instruct \cite{qwen2audio}. We emulate standard zero-shot deployment practices, 
applying extensive prompt engineering techniques, including direct numerical scoring, 
detailed rubric guidance, and Chain-of-Thought (CoT) \cite{chainofthought} reasoning, 
we attempt to prompt the base model to evaluate synthetic speech samples. However, zero-shot 
prompting yields severe operational failures: the model exhibits high score compression and 
fails to achieve acceptable correlation with human ground-truth ratings. This demonstrates 
that out-of-the-box Audio-LLMs cannot be reliably deployed as zero-shot release blockers in 
production pipelines.

Rather than discarding the Audio-LLM, we investigate the perceptual 
root cause of this failure. By analyzing the human-curated SOMOS dataset \citep{Maniati_2022}, 
we extract the objective acoustic properties, specifically speech flow and pitch variability, 
that most strongly drive human MOS perception. Using these, we construct a 
diagnostic benchmark to test whether Qwen2-Audio-Instruct can perceive basic, low-level 
speech characteristics. Our benchmark reveals that the base model struggles to correctly 
categorize basic attributes such as slow/fast tempo or flat/expressive intonation, 
establishing that its high-level MOS failure may stem from a fundamental lack of 
low-level acoustic feature recognition.

To resolve these perceptual blind spots, we propose a hybrid system design that uses SOMOS 
solely as the alignment mechanism to train a fast, offline LLM evaluator. By applying 
Low-Rank Adaptation (LoRA) \citep{lora}, we fine-tune the base model directly on the task of 
human MOS prediction. This parameter-efficient adaptation restores strong correlation with 
human judgment while maintaining a low computational paradigm required for fast offline batch 
processing. 

Our main contributions are as follows: \textbf{First}, we validate that out-of-the-box 
zero-shot prompting of Qwen2-Audio-7B-Instruct fails to predict speech MOS scores
that are aligned with human scores, hence a more sophisticated approach is needed. 
\textbf{Second}, we analyze the SOMOS dataset to map objective speech attributes to human MOS 
perception, establishing a diagnostic unit-testing benchmark that exposes the baseline model's 
low-level perceptual blind spots. For this purpose, we also define a novel metric called 
\textit{expressivity}, that encapsulates the human perception of expressiveness. \textbf{Third}, 
we implement a parameter-efficient LoRA fine-tuning strategy that explicitly aligns the 
Audio-LLM with human MOS, providing teams with an accurate, low-cost blueprint for 
automated offline speech evaluation.

\section{Related Work}

\subsection{Traditional MOS prediction systems}

Speech MOS evaluation, is a standardized method for rating the 
perceived quality, clarity, and naturalness of audio or speech. Human listeners 
score speech samples on a numeric scale from 1 (bad) to 5 (excellent), and the 
ratings are averaged to yield a final score. Historically, automating speech 
quality evaluation in production pipelines relied on specialized deep neural 
networks trained to map audio features directly to human MOS. 
Models such as UTMOS \citep{utmos} and NISQA \citep{Mittag_2021} established the 
baseline for predicting overall naturalness and multidimensional signal 
degradation, while incorporating prosodic and linguistic features improved performance 
\cite{Vioni_2023}. Furthermore, the introduction of large-scale, TTS-specific datasets 
like SOMOS \citep{Maniati_2022} provided the necessary human ground-truth for evaluating 
neural speech synthesis. However, these systems cannot fully capture the variability of 
speech and often lack robustness when applied to out-of-domain data \cite{generalizationabilitymosprediction}. 
Although computationally lightweight, they output a single numerical score with limited 
diagnostic value. As speech systems become more sophisticated, their evaluation requires 
models that can adapt to diverse speech characteristics and can not only output a single score 
but also isolate the underlying acoustic failures and provide an explanatory 
evaluation of the quality.

\subsection{Large Audio-Language Models}

The recent paradigm shift toward multimodal foundation models has enabled 
Large Language Models (LLMs) to process audio natively. Architectures such as 
Qwen2-Audio \cite{qwen2audio}, Audio Flamingo \cite{audioflamingo} and SALMONN \cite{salmonn} 
integrate audio encoders directly with large autoregressive decoders, achieving 
state-of-the-art performance on speech recognition, translation, and audio-based instruction 
following. Despite these semantic and conversational capabilities, applying these 
models to rigorous acoustic quality evaluation remains challenging. 
Our diagnostic benchmarking demonstrates that robust semantic processing 
does not inherently guarantee the perception of low-level acoustic primitives, 
such as audio artifacts, speech flow and pitch expressivity, which are critical for 
assessing the quality of a speech sample generated by a TTS system.

\subsection{Using Audio-LLMs for Speech Evaluation}

In recent years, researchers have increasingly investigated 
using Audio-LLMs directly for speech quality evaluation and MOS prediction \cite{wang2025enablingauditorylargelanguage}. 
Recent frameworks have focused on transforming these models into descriptive 
evaluators capable of generating natural language rationales for audio degradation. 
For instance, \cite{chen2025} introduced Alignment with LLM Distillation (ALLD), 
employing complex token-level distillation from expert models to teach Audio-LLMs 
multidimensional evaluation. Similarly, the SpeechLLM-as-Judges paradigm 
\cite{wang2026} utilizes Chain-of-Thought (CoT) reasoning and reward optimization 
to produce structured, explanation-based quality assessments. However, these approaches 
rely on computationally intensive distillation, multi-step reasoning, and large-scale 
models, resulting in substantial resource requirements that limit their 
suitability for real-time or resource-constrained applications.

\begin{figure}[t]
    \centering
    \includegraphics[width=0.35\textwidth]{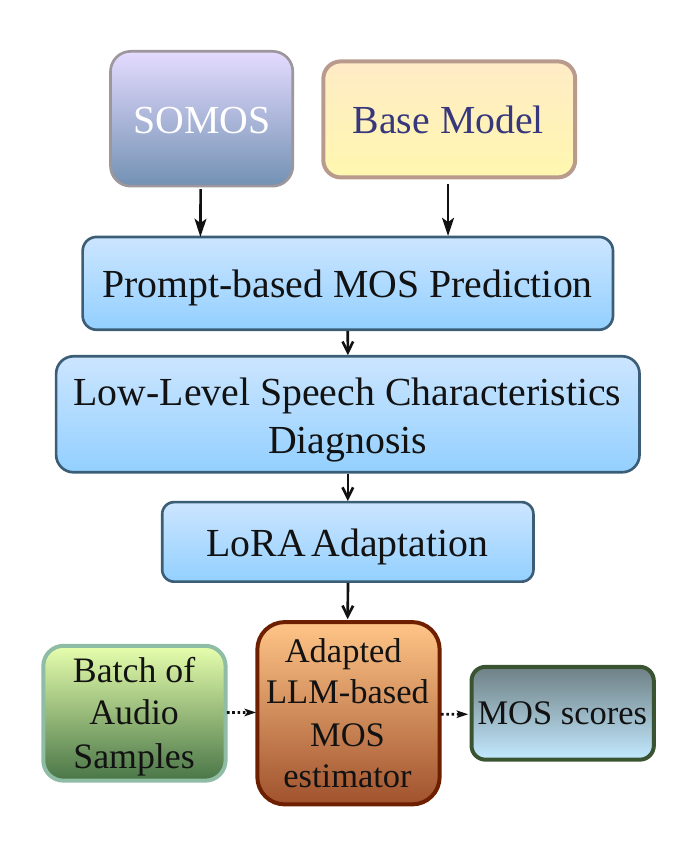}
    \caption{Proposed framework design for adapting a base Audio LLM into a MOS prediction system.}
    \label{fig:system}
\end{figure}

\section{System Design}

The proposed framework, as depicted in Figure \ref{fig:system}, is a pipeline with the
purpose of adapting a base Audio-LLM into an offline speech naturalness estimator.
The objective is to provide an automatic offline pre-screening signal that
can support faster comparison of candidate systems and help prioritize samples for possible
future human evaluation. Our framework provides a blueprint that guides teams through three 
distinct phases: zero-shot probing, diagnostic evaluation and parameter-efficient alignment.  
At inference time, the system receives batches of audio samples generated by a candidate TTS model,
together with a fixed instruction prompt and returns a direct scalar MOS value. 
Operating in a 4-bit quantized Qwen2-Audio-7B-Instruct, augmented with LoRA weights,
it minimizes GPU memory footprints, allowing large batch sizes during offline evaluation.
The system can be used as a fast sorting layer, where extremely low-scoring candidates
are immediately flagged as regression failures and blocked from deployment without
consuming human review budget.

\section{Zero-Shot Audit}

\subsubsection{Prompt Strategies}

Using Qwen2-Audio-7B-Instruct as a representative state-of-the-art open-source Audio-LLM, 
we perform a baseline feasibility assessment, probing across four prompting paradigms to establish whether the 
base model can comprehend speech quality evaluation and generate plausible numerical scores 
given the quality of the speech sample, without further finetuning.

Initially, we evaluated direct instructions, prompting the model to output a scalar score 
reflecting the overall quality and naturalness of the speech sample. Prompt variations 
incorporated explicit MOS definitions, contextual framing identifying samples as synthetic TTS 
outputs, strict scoring guidelines, and targeted dimensions focusing on prosody, expressiveness, 
emotional range, and intelligibility. In each scenario, the model failed to generate scores 
aligned with the perceptual quality of the sample, evaluating transcription accuracy rather than acoustic fidelity or prosodic naturalness, 
resulting in consistently high scores ($> 4.5$).

Continuing, we explored a few-shot in-context learning strategy by conditioning the model 
on $25$ examples consisting of \{audio, ground-truth MOS\} pairs and then prompting to 
evaluate a new sample. What we found is that the model simply replicated the scalar values 
present in the prompt context rather than evaluating the perceptual attributes of the 
target audio. 

To examine whether the model could judge relative quality without requiring continuous scalar 
regression, we formulated a pairwise preference comparison task. We created pairs of speech 
samples, along with a true answer 'A' or 'B', corresponding to which sample has higher MOS 
(``winner''). This configuration revealed a severe, deterministic positional bias, 
wherein the model systematically selected the first option (Sample A) regardless of sample 
ordering or ground-truth quality disparities. 

Finally, we implemented a Chain-of-Thought (CoT) rationale-first approach, requiring the 
model to generate qualitative evaluations across specific dimensions, such as intelligibility, 
prosody, naturalness, technical fidelity, and listening effort, prior to outputting a final 
rating. While the model occasionally identified concrete acoustic defects such as robotic voice 
or digital artifacts in its text analysis, its subsequent numerical scores remained inflated, 
contradicting its own analysis. This aligned with our instinct that the model might be capable 
of recognizing failure points in the speech sample but can't translate them into a numerical 
score.

\subsubsection{Mapping text to numerical score via LLM}

We test this theory using a two-stage decoupled pipeline. In the first stage, our base Audio-LLM 
generates qualitative textual descriptions across predefined acoustic dimensions without 
emitting scalar ratings. In the second stage, these qualitative evaluations are passed to a 
text-only LLM (Llama-3-8B-Instruct \cite{llama3}) tasked solely with mapping the diagnostic text to 
numerical MOS values. Experimenting with different generation temperatures for Qwen2-Audio-7B-Instruct, 
we observe minor improvements on the Pearson correlation between the predicted scores and the 
true scores, as seen in Table \ref{tab:descr}. We conclude that the base model is, at some level, 
able to understand the quality of the speech sample but struggles to map this understanding to a 
numerical score. This demonstrates that separating qualitative diagnostic generation from 
scalar scoring provides a practical, zero-shot mitigation strategy for speech evaluation 
pipelines prior to model fine-tuning.

\begin{table}[htbp]
\centering
\renewcommand{\arraystretch}{0.9}
\resizebox{0.5\linewidth}{!}{
\begin{tabular}{lc}
\hline
\textbf{Temperature} & \textbf{Pearson (r)}\\
\hline
Base & -0.0312 \\
\hline
0.5 & 0.1603 \\
1.0 & 0.1681 \\
1.5 & 0.2467 \\
2.0 & 0.0422 \\\hline
\end{tabular}
}
\caption{Pearson (r) correlation between the generated scores and ground-truth MOS values, 
across different temperatures.}
\label{tab:descr}
\end{table}

\section{Diagnostic Benchmarking}

\subsection{SOMOS Dataset Analysis}

In order to improve the model's ability to evaluate a speech sample and thus integrate it
into a real-world quality evaluation system, we need to understand what aspects of speech are 
important during the process of evaluation. We analyzed the SOMOS dataset, extracting acoustic 
properties from the samples and categorizing them into several dimensions. 

\subsubsection{Dataset}

Samsung Open MOS Dataset (SOMOS) is a large-scale MOS dataset consisting solely 
of neural text-to-speech samples. It comprises of 20K synthetic utterances of the 
LJ Speech voice generated from 200 different TTS systems.
MOS naturalness evaluations are collected via crowdsourcing on Amazon Mechanical Turk.

\subsubsection{Speech Flow}

\begin{figure}
    \centering
    \includegraphics[width=0.48\textwidth]{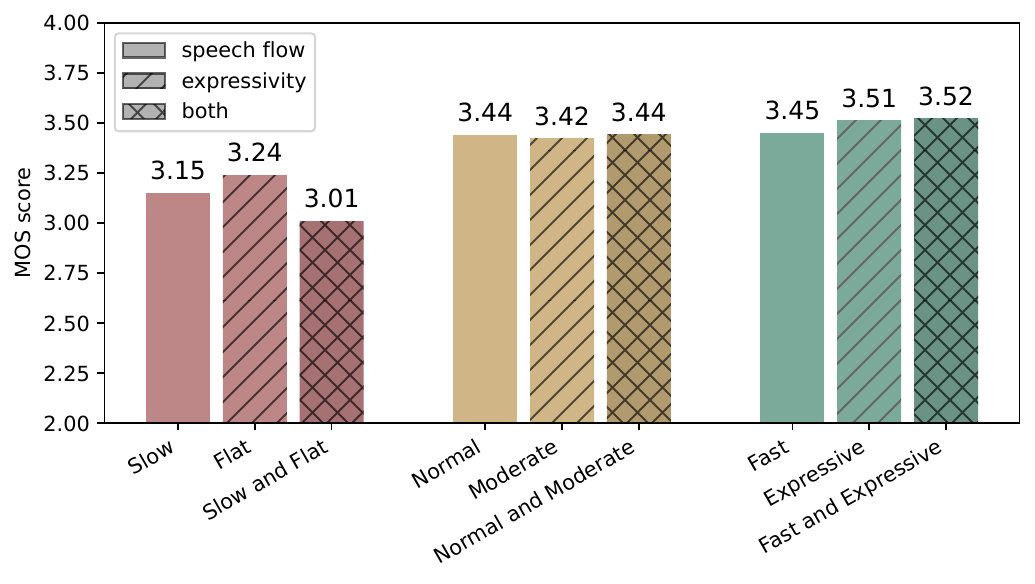}
    \caption{Average MOS score for each characteristic.}
    \label{fig:metrics}
\end{figure}

Speech flow describes the pace of the utterance, as well as the number and 
duration of temporary breaks in speaking. It defines the mood, emotion and state of the speaker 
and consequently affects the naturalness of speech. To analyze whether speaking rate influences 
human evaluation, we categorized dataset samples into discrete tiers (slow, normal, and fast) 
using the sentence length to duration ratio as a proxy for tempo.
Through experimentation and empirical examination of the metric distribution, we found that 
samples with a ratio below $12$ exhibited slow delivery with extended inter-word pauses, 
whereas ratios exceeding $20$ corresponded to rapid, unpaused speech. 
As depicted in Figure \ref{fig:metrics}, samples categorized as slow received a slightly lower 
mean MOS compared to their normal and fast counterparts, indicating a possible incline of the 
human raters to evaluate the samples based on the speech flow of the speaker.

\begin{table*}[t]
\centering
\begin{tabular}{lcccccc}
\toprule
 & \multicolumn{3}{c}{Main} & \multicolumn{3}{c}{Easy} \\
\cmidrule(lr){2-4} \cmidrule(lr){5-7}
\multicolumn{1}{c}{Model} & Speed & Expressive & Content & Speed & Expressive & Content \\
\midrule
Qwen2-Audio-7B-Instruct & 21.9\% & 26.5\% & 29.5\% & 33\% & 33.3\% & 58.2\% \\
Audio-Flamingo3 & 16.9\% & 27.4\% & 36.5\% & 39.8\% & 37.3\% & 42.3\% \\
\bottomrule
\end{tabular}
\caption{Performance of each model on the benchmark, evaluating speech rate, expressiveness and 
content perception. The easy mode considers a 2-class classification.}
\label{tab:bench}
\end{table*}

\subsubsection{Pitch Variability}

One other important aspect of human speech is the expressiveness, which can be attributed to 
pitch modulations. High deviation and frequent, dynamic alternations from the mean frequency indicate 
expressive, melodic speech, whereas low variations are often associated with a ``flat'' speech.
To quantify expressiveness, we extracted fundamental pitch dynamics from the signal: 
mean fundamental frequency ($F_0$), pitch standard deviation, pitch average baseline crossing 
rate, and average pitch velocity and we min-max normalized the values across the dataset. 
We formulated a composite \textit{expressivity} \eqref{eq:expressivity} metric as a weighted 
linear combination of the standard deviation, mean crossing rate, and velocity, optimizing the 
feature weights via grid search (step size $0.01$) to maximize correlation with ground-truth human MOS. 

\begin{equation}\label{eq:expressivity}
  \text{expressivity} = (0.46\text{sd} + 0.05\text{mcr} + 0.49\text{vel})
\end{equation}

Based on the resulting metric distribution, samples were discretized into three expressivity 
tiers: flat ($< 0.15$), moderate ($0.15-0.35$), and expressive ($> 0.35$). As illustrated 
in Figure \ref{fig:metrics}, flat speech exhibits a slightly lower average human MOS, 
whereas expressive speech receives higher quality ratings. This confirms that human 
evaluators penalize monotonous, unexpressive intonation, establishing pitch expressivity 
as a critical factor in determining the quality of a speech sample.

\subsection{Benchmark}

\subsubsection{Setup}

Leveraging these discretized acoustic labels, we construct a diagnostic benchmark suite designed 
to measure an Audio-LLM's low-level perceptual abilities across three core dimensions: speech flow, 
expressiveness and content alignment. We frame the benchmark as a balanced 3-class 
multiple-choice classification task. The model is provided with audio inputs and a 
targeted question, and its performance is measured by its accuracy in outputting the 
correct categorical label. Sampling from the SOMOS dataset, we retained utterances between 
$2$ and $20$ seconds in duration, and created three distinct sub-datasets of 600 samples each 
($n=200$ per class):

\begin{itemize}
  \item Speech Flow: Probes tempo perception by requiring the model to classify single clips 
  into \textit{slow}, \textit{normal}, or \textit{fast}.
  \item Expressivity: Examines intonation and expressiveness sensitivity by tasking the model 
  to categorize single clips into \textit{flat}, \textit{moderate}, or \textit{expressive}.
  \item Content Alignment: Tests cross-clip phonetic and linguistic discrimination ability by 
  presenting audio pairs and classifying them as identical audio (\textit{same}), identical text 
  with a different utterance (\textit{utterance}), or completely distinct samples (\textit{different}).
\end{itemize}

\subsubsection{Evaluation}

We evaluated Qwen2-Audio-7B-Instruct and Audio Flamingo 3, which is built similarly with the 
former using Whisper as the audio encoder and Qwen2.5-7B as the LLM backbone. In Table \ref{tab:bench}, 
we demonstrate the macro averaged F1-score across all categories for each model, because the 
accuracy metric alone is insufficient, due to repetitive answers or biases. For each dataset, 
we also evaluate an ``easy mode'', which is the same dataset without the middle category 
(normal, moderate and utterance), potentially imposing a less challenging task for the model. 

For both models we used temperature = $0.7$. We observe that both models struggle with low level 
speech attributes. Qwen particularly doesn't output the labels ``slow'', ``flat'' or ``utterance''. 
This indicates that these models have not been trained to perceive audio characteristics, that 
are valuable when assessing the naturalness of a speech sample, imposing a restrictive reasoning 
space that factors in basic speech characteristics for the evaluation. 

\section{Parameter-Efficient Domain Adaptation}

As scalar quality scoring falls outside the zero-shot capabilities of base Audio-LLMs, 
integrating them in a system requires targeted adaptation. In this section, we 
explore whether Low-Rank Adaptation (LoRA) can efficiently calibrate Qwen2-Audio-7B-Instruct 
for automated MOS estimation under strict data and compute constraints 
(4-bit quantization, single GPU). We focus our investigation on two critical data engineering 
decisions: mitigation of dataset score skew and target output formatting.

\begin{table}
\centering
\resizebox{0.95\linewidth}{!}{
\begin{tabular}{l|ccc}
\toprule
\textbf{mode + balance\_ratio} & \textbf{LoRA} & \textbf{r}$_{pearson}$ & SRCC \\
\midrule
textual + 0.8 & r = 16, a = 16 & 0.0406 & 0.0287 \\
\hline
numerical + 0.8 & r = 16, a = 16 & 0.6332 & 0.6438 \\
numerical + 1.0 & r = 16, a = 16 & 0.5517 & 0.5971 \\
numerical + 1.0 &  r = 16, a = 32 & 0.5435 & 0.5986 \\
numerical + 0.0 &  r = 16, a = 16 & 0.6371 & 0.6426 \\
numerical + 0.0 &  r = 16, a = 32 & \textbf{0.6506} & \textbf{0.6582} \\
\bottomrule
Base\_Model &  & 0.0028 & 0.02 \\
\hline
UTMOSv2 &  & 0.4310 & 0.4193
\end{tabular}
}
\caption{Pearson (r) and Spearman (SRCC) correlation between the utterance-level predictions and ground truth MOS scores 
across each different finetuning setting. \textbf{Bold} indicates the best result.}
\label{tab:lora}
\end{table}

\subsection{Dataset}

We utilize the SOMOS training set to construct our instruction-tuning dataset. The dataset 
exhibits an unbalanced score distribution, with most samples lying in the $3.0$--$4.0$ range.
To prevent the model from developing a majority-class prediction bias during fine-tuning, 
we explore a systematic bin-balancing strategy. 
We partition the distribution into four discrete bins: $[1.0, 2.0)$, $[2.0, 3.0)$, $[3.0, 4.0)$, and $[4.0, 5.0]$. 
We define a \textit{balance ratio} $\beta \in [0.0, 1.0]$ that downsamples overrepresented 
bins using the second-lowest population bin as an anchor. Under this formulation, 
$\beta = 1.0$ yields a perfectly uniform class distribution at the cost of total sample volume, 
whereas $\beta = 0.0$ preserves the original full-volume dataset.

The training dataset follows an instruction-tuning format, with each training instance 
formatted as an instruction triplet: $\{x_{\text{audio}}, x_{\text{prompt}}, y_{\text{target}}\}$. 
The input prompt $x_{\text{prompt}}$ is standardized across experiments: 

\begin{quote}
\small\texttt{"Listen to the audio speech sample and evaluate it based on the naturalness of the speech and the audio quality, on a scale of 1(bad) to 5(excellent). The number can have decimals."}
\end{quote}

To identify the optimal target representation for autoregressive language models performing 
regression, we evaluate two target formatting modes: (1) \textbf{Direct Scalar Mode:} 
The ground-truth $y_{\text{target}}$ consists strictly of the numerical float 
(e.g., \texttt{"3.29"}), (2) \textbf{Textual Mode:} The numerical score is embedded inside a 
conversational template (e.g., \texttt{"I would evaluate the sample with 3.29"}), sampled 
dynamically from a pre-defined template pool during dataset construction. The ground-truth MOS
scores were rounded to two decimal places to promote stable convergence during finetuning and 
reduce unnecessary target variance.

\subsection{Results \& Discussion}

All experiments were conducted on Qwen2-Audio-7B-Instruct quantized to 
4-bit precision, with a batch size of 2, learning rate to $1e-4$, 
trained for a single epoch. The model was tested on the test set of SOMOS. The results are 
summarized in the Table \ref{tab:lora}.

For the first experiment, we finetuned the model using the ``textual'' mode and 0.8 balance ratio. 
This resulted in almost zero correlation, therefore we chose to finetune the model in a 
direct scalar manner. From the results we can draw the following conclusions: (1) The textual 
mode fails because textual tokens far outnumber the numerical score tokens, the optimizer rapidly 
minimizes loss by predicting text correctly while assigning arbitrary probability distributions 
to the scalar output tokens. (2) Contrary to the hypothesis that score skew would induce 
numerical bias, balancing the dataset degraded correlation performance. This means that
data volume outweighs class imbalance for low-rank regression alignment. (3) Expanding the 
LoRA rank ($r$) or scaling factor ($\alpha$) yielded minimal returns on correlation 
metrics. Model capacity is not the bottleneck in domain adaptation but rather, key projection 
layers across the audio-language bridge (4) Finetuning on the whole training set with $0.53\%$ of
the model's parameters, obtains $0.65$ Pearson correlation against a zero correlation of the base model, 
even surpassing UTMOS. This validates that parameter-efficient adaptation provides an accurate, 
low-overhead blueprint for deploying automated quality gates in production speech pipelines.

\section{Conclusion}

In this work, we present a practical framework for adapting an Audio-LLM into an
automated MOS evaluation system in speech production pipelines. 
We examined whether off-the-shelf Qwen2-Audio-7B-Instruct is capable to map speech quality to a 
numerical score through prompting, and designed a benchmark for evaluating the capacity of the model
to comprehend low level speech properties. Finally, we introduced a parameter-efficient LoRA 
strategy that successfully adjusts the model for MOS prediction aligned with human perception. 
The resulting pipeline operates as an auto-filtering synthetic speech system at a fraction of the 
cost and latency of crowdsourced panels. This work paves the way for utilizing Audio-LLMs as 
reliable and versatile evaluators in speech generation systems. 

\section{Limitations}

While our proposed approach achieves significant improvement in lightweight MOS prediction, 
there are several limitations worth mentioning. First, our observations are based on in-domain
evaluation, highlighting the need to improve coverage across out-of-domain datasets. 
Second, SOMOS is an English synthetic speech dataset, therefore our findings regarding 
low-level acoustic probing and LoRA adaptation may not immediately transfer to 
multi-lingual TTS systems. Third, all experiments are conducted primarily on Qwen2-Audio-7B-Instruct 
as a representative open-source model. While our methodology and diagnostic framework are 
model-agnostic, validating these findings across other audio-language architectures remains an 
avenue for future work.

\section*{Ethics Statement}

Rather than replacing entirely human evaluators, the proposed automated system complements human 
panels by filtering obvious acoustic failures and clear successes, allowing teams to 
concentrate crowdsourced human rating budgets on samples of ambiguous quality.
To mitigate the environmental impact of large multimodal AI, our framework prioritizes 
parameter-efficient adaptation over full-parameter fine-tuning, reducing GPU memory footprint.

\bibliography{custom}

@inproceedings{Maniati_2022,
   title={SOMOS: The Samsung Open MOS Dataset for the Evaluation of Neural Text-to-Speech Synthesis},
   url={http://dx.doi.org/10.21437/Interspeech.2022-10922},
   DOI={10.21437/interspeech.2022-10922},
   booktitle={Interspeech 2022},
   publisher={ISCA},
   author={Maniati, Georgia and Vioni, Alexandra and Ellinas, Nikolaos and Nikitaras, Karolos and Klapsas, Konstantinos and Sung, June Sig and Jho, Gunu and Chalamandaris, Aimilios and Tsiakoulis, Pirros},
   year={2022},
   pages={2388--2392},
}

@inproceedings{utmos,
  title     = {The T05 System for The {V}oice{MOS} {C}hallenge 2024: Transfer Learning from Deep Image Classifier to Naturalness {MOS} Prediction of High-Quality Synthetic Speech},
  author    = {Baba, Kaito and Nakata, Wataru and Saito, Yuki and Saruwatari, Hiroshi},
  booktitle = {IEEE Spoken Language Technology Workshop (SLT)},
  year      = {2024},
  pages     = {818--824},
  doi       = {10.1109/SLT61566.2024.10832315},
}

@inproceedings{Mittag_2021,
   title={NISQA: A Deep CNN-Self-Attention Model for Multidimensional Speech Quality Prediction with Crowdsourced Datasets},
   url={http://dx.doi.org/10.21437/Interspeech.2021-299},
   DOI={10.21437/interspeech.2021-299},
   booktitle={Interspeech 2021},
   publisher={ISCA},
   author={Mittag, Gabriel and Naderi, Babak and Chehadi, Assmaa and Möller, Sebastian},
   year={2021},
   month=Aug, pages={2127--2131},
}

@misc{qwen3tts,
      title={Qwen3-TTS Technical Report}, 
      author={Hangrui Hu and Xinfa Zhu and Ting He and Dake Guo and Bin Zhang and Xiong Wang and Zhifang Guo and Ziyue Jiang and Hongkun Hao and Zishan Guo and Xinyu Zhang and Pei Zhang and Baosong Yang and Jin Xu and Jingren Zhou and Junyang Lin},
      year={2026},
      eprint={2601.15621},
      archivePrefix={arXiv},
      primaryClass={cs.SD},
      url={https://arxiv.org/abs/2601.15621}, 
}

@misc{fishaudios2,
      title={Fish Audio S2 Technical Report}, 
      author={Shijia Liao and Yuxuan Wang and Songting Liu and Yifan Cheng and Ruoyi Zhang and Tianyu Li and Shidong Li and Yisheng Zheng and Xingwei Liu and Qingzheng Wang and Zhizhuo Zhou and Jiahua Liu and Xin Chen and Dawei Han},
      year={2026},
      eprint={2603.08823},
      archivePrefix={arXiv},
      primaryClass={cs.SD},
      url={https://arxiv.org/abs/2603.08823}, 
}

@misc{qwen2audio,
      title={Qwen2-Audio Technical Report}, 
      author={Yunfei Chu and Jin Xu and Qian Yang and Haojie Wei and Xipin Wei and Zhifang Guo and Yichong Leng and Yuanjun Lv and Jinzheng He and Junyang Lin and Chang Zhou and Jingren Zhou},
      year={2024},
      eprint={2407.10759},
      archivePrefix={arXiv},
      primaryClass={eess.AS},
      url={https://arxiv.org/abs/2407.10759}, 
}

@misc{chainofthought,
      title={Chain-of-Thought Prompting Elicits Reasoning in Large Language Models}, 
      author={Jason Wei and Xuezhi Wang and Dale Schuurmans and Maarten Bosma and Brian Ichter and Fei Xia and Ed Chi and Quoc Le and Denny Zhou},
      year={2023},
      eprint={2201.11903},
      archivePrefix={arXiv},
      primaryClass={cs.CL},
      url={https://arxiv.org/abs/2201.11903}, 
}

@misc{lora,
      title={LoRA: Low-Rank Adaptation of Large Language Models}, 
      author={Edward J. Hu and Yelong Shen and Phillip Wallis and Zeyuan Allen-Zhu and Yuanzhi Li and Shean Wang and Lu Wang and Weizhu Chen},
      year={2021},
      eprint={2106.09685},
      archivePrefix={arXiv},
      primaryClass={cs.CL},
      url={https://arxiv.org/abs/2106.09685}, 
}

@misc{audioflamingo,
      title={Audio Flamingo 3: Advancing Audio Intelligence with Fully Open Large Audio Language Models}, 
      author={Arushi Goel and Sreyan Ghosh and Jaehyeon Kim and Sonal Kumar and Zhifeng Kong and Sang-gil Lee and Chao-Han Huck Yang and Ramani Duraiswami and Dinesh Manocha and Rafael Valle and Bryan Catanzaro},
      year={2025},
      eprint={2507.08128},
      archivePrefix={arXiv},
      primaryClass={cs.SD},
      url={https://arxiv.org/abs/2507.08128}, 
}

@misc{salmonn,
      title={SALMONN: Towards Generic Hearing Abilities for Large Language Models}, 
      author={Changli Tang and Wenyi Yu and Guangzhi Sun and Xianzhao Chen and Tian Tan and Wei Li and Lu Lu and Zejun Ma and Chao Zhang},
      year={2024},
      eprint={2310.13289},
      archivePrefix={arXiv},
      primaryClass={cs.SD},
      url={https://arxiv.org/abs/2310.13289}, 
}

@misc{chen2025,
      title={Audio Large Language Models Can Be Descriptive Speech Quality Evaluators}, 
      author={Chen Chen and Yuchen Hu and Siyin Wang and Helin Wang and Zhehuai Chen and Chao Zhang and Chao-Han Huck Yang and Eng Siong Chng},
      year={2025},
      eprint={2501.17202},
      archivePrefix={arXiv},
      primaryClass={cs.SD},
      url={https://arxiv.org/abs/2501.17202}, 
}

@misc{wang2026,
      title={SpeechLLM-as-Judges: Towards General and Interpretable Speech Quality Evaluation}, 
      author={Hui Wang and Jinghua Zhao and Yifan Yang and Shujie Liu and Junyang Chen and Yanzhe Zhang and Shiwan Zhao and Jinyu Li and Jiaming Zhou and Haoqin Sun and Yan Lu and Yong Qin},
      year={2026},
      eprint={2510.14664},
      archivePrefix={arXiv},
      primaryClass={cs.SD},
      url={https://arxiv.org/abs/2510.14664}, 
}

@misc{llama3,
  title={The Llama 3 Herd of Models},
  author={Dubey, Abhimanyu and Jauhri, Abhinav and Pandey, Abhinav and Kadian, Abhishek and Al-Dahle, Ahmad and others},
  year={2024},
  url={https://arxiv.org/abs/2407.21783}
}

@misc{generalizationabilitymosprediction,
      title={Generalization Ability of MOS Prediction Networks}, 
      author={Erica Cooper and Wen-Chin Huang and Tomoki Toda and Junichi Yamagishi},
      year={2022},
      eprint={2110.02635},
      archivePrefix={arXiv},
      primaryClass={eess.AS},
      url={https://arxiv.org/abs/2110.02635}, 
}

@inproceedings{Vioni_2023,
   title={Investigating Content-Aware Neural Text-to-Speech MOS Prediction Using Prosodic and Linguistic Features},
   url={http://dx.doi.org/10.1109/ICASSP49357.2023.10096255},
   DOI={10.1109/icassp49357.2023.10096255},
   booktitle={ICASSP 2023 - 2023 IEEE International Conference on Acoustics, Speech and Signal Processing (ICASSP)},
   publisher={IEEE},
   author={Vioni, Alexandra and Maniati, Georgia and Ellinas, Nikolaos and Sig Sung, June and Hwang, Inchul and Chalamandaris, Aimilios and Tsiakoulis, Pirros},
   year={2023}}

@misc{wang2025enablingauditorylargelanguage,
      title={Enabling Auditory Large Language Models for Automatic Speech Quality Evaluation}, 
      author={Siyin Wang and Wenyi Yu and Yudong Yang and Changli Tang and Yixuan Li and Jimin Zhuang and Xianzhao Chen and Xiaohai Tian and Jun Zhang and Guangzhi Sun and Lu Lu and Yuxuan Wang and Chao Zhang},
      year={2025},
      eprint={2409.16644},
      archivePrefix={arXiv},
      primaryClass={eess.AS},
      url={https://arxiv.org/abs/2409.16644}, 
}
\bibliographystyle{acl_natbib}

\appendix


\end{document}